\documentstyle[11pt]{article}

\newcommand{\half}{{\scriptstyle{\frac{1}{2}}}}

\newcommand{\del}{{\mbox{\boldmath $\nabla$}}}
\newcommand{\BE}{\begin{equation}}
\newcommand{\EE}{\end{equation}}
\newcommand{\BA}{\begin{eqnarray}}
\newcommand{\EA}{\end{eqnarray}}

\begin{document}

\begin{titlepage}

\vspace*{1mm}
\begin{center}

            {\LARGE{\bf Newtonian gravity from \\ 
                 Higgs condensates  }}

\vspace*{11mm}
{\Large  M. Consoli }
\vspace*{3mm}\\
{\large
Istituto Nazionale di Fisica Nucleare, Sezione di Catania \\
Corso Italia 57, 95129 Catania, Italy}
\vspace*{3mm}\\
and \\
\vspace*{3mm}
{\Large F. Siringo}
\vspace*{2.5mm}\\
{\large Dipartimento di Fisica dell' Universit\`a di Catania \\
Corso Italia 57, 95129 Catania, Italy}
\vspace{7mm}\\
\end{center}
\begin{center}
{\bf Abstract}
\end{center}

We propose a description of {\it Newtonian } gravity
as a long wavelength excitation of the scalar condensate inducing 
electroweak symmetry breaking. Indeed, one finds a
$-{{G_F}\over{\eta}}{{m_im_j}\over{r}}$ long-range potential
where $G_F$ is the Fermi constant and 
$\eta\equiv {{M^2_h}\over{2m^2}} $ is determined by the ratio between the Higgs 
mass $M_h$ and the mass $m$ of the elementary quanta of the symmetric phase
(`phions'). The parameter $\eta$ would 
diverge in a true continuum theory so that its magnitude represents a measure of
non-locality of the underlying field theory. 
By identifying $G\equiv {{G_F}\over{\eta}}$ with the Newton 
constant and assuming the range of Higgs mass 
$M_h \sim 10^{2}-10^{3}$ GeV one obtains $m=10^{-4}-10^{-5}$ eV and
predicts typical
`fifth-force' deviations below the centimeter scale. 
Relation to Einstein gravity and string theory is discussed.
The crucial role of the first-order 
nature of the phase transition for the solution of the so-called 
`hierarchy problem' is emphasized. The possible relevance of the picture for
the self-similarity of the universe and for a new approach to
the problem of dark matter is discussed.

\end{titlepage}
 
\setcounter{page}{1}

\vfill
\eject
\setcounter{equation}{0}
\section{Introduction}

In this paper we shall present a simple mechanism to explain the physical
origin of {\it Newtonian} gravity. However, the motivations for our proposal 
depend crucially on the description of spontaneous symmetry breaking (SSB)
in scalar quantum field theories.
For this reason, we shall address the main issue in sects.2 and 3, 
after the preliminary discussion presented in this Introduction.

The `condensation' of a scalar field, i.e. the transition 
from a symmetric phase where $\langle \Phi \rangle=0 $
to the physical vacuum where
$\langle \Phi \rangle \neq 0  $, has been traditionally described as an
essentially classical phenomenon (with perturbative quantum 
corrections). In this picture, one uses a classical potential 
\BE
\label{clpot}
    V_{\rm cl}(\phi)= {{1}\over{2}} m^2 \phi^2 + 
{{\lambda}\over{4!}}\phi^4  
\EE
where the phase transition, as one varies the $m^2$ parameter, is 
second order and occurs at $m^2=0$. 

As discussed in ref.\cite{mech},  the 
question of vacuum stability is more subtle in the quantum theory. Here, 
the starting point is the Hamiltonian operator
\BE 
H = \, : \! \int \! d^3 x \left[ \frac{1}{2} \left( \Pi^2 + 
(\del \Phi)^2 + m^2 \Phi^2 \right) + 
\frac{\lambda}{4!} \Phi^4 \right] \! : \; 
\EE
after quantizing the scalar field $\Phi$ and the canonical momentum $\Pi$
 in terms of annihilation and creation operators
$a_{\bf k}$, $a^{\dagger}_{\bf k}$
 of a reference vacuum state $|o\rangle$
($a_{\bf k}|o\rangle=\langle o|a^{\dagger}_{\bf k}$=0).
These satisfy the commutation relations 
\BE
[ a_{\bf k}, a^{\dagger}_{{\bf k}'} ] = \delta_{{\bf k},{\bf k}'}.
\EE
and, due to normal ordering,
the quadratic part of the Hamiltonian has the usual form 
($E_k=\sqrt{ k^2 + m^2}$ )
\BE
H_2 = \sum_{\bf k} E_k a^{\dagger}_{\bf k} a_{\bf k}.
\EE
for the elementary quanta of the symmetric phase (`phions'). 

Now the trivial vacuum $|o\rangle$
 where $\langle\Phi\rangle=0$ is clearly locally 
stable if phions have a physical mass $m^2 >0$.
  However, is an $m^2>0$ 
symmetric vacuum necessarily {\it globally} stable ?  Could the phase 
transition actually be first order, occurring at some small but positive 
value of the physical mass squared
$m^2 > 0$? The question is not entirely trivial just because \cite{mech} 
the static limit of the
2-body phion-phion interaction is not always repulsive. Besides the
tree-level repulsive potential
there is an induced 
attraction from higher-order graphs. In this case, 
for sufficiently small values of $m$, the trivial `empty' state  $|o\rangle$ 
may not be the physical vacuum.

  The answer to the question depends on the form of the 
{\it effective potential} $V_{\rm eff}(\phi)$ 
and it is not surprising that different approximations
may lead to contradictory results on this crucial issue. The situation is 
similar to the Bose-Einstein condensation in condensed matter
that is a first-order 
phase transition in an ideal gas. However, 
in interacting systems the issue is more delicate and often difficult to be
settled experimentally. Theoretically is predicted to be 
a second-order transition in some approximations but it may appear as
a weak first-order transition in other approximations \cite{trento}. 

We shall refer to \cite{mech,zeit} for details on the 
structure and the meaning of  various types of
approximations to the effective
potential and just report a
few basic results:

\par~~~~i) the phase transition 
 is indeed first order as in the case of the simple one-loop potential. This
is easy to realize 
if one performs a variational procedure, within a simple
class of trial states that includes $|o\rangle$. In this case, one
finds \cite{cian} that the $m=0$ theory lies in the broken phase. Therefore
the phase transition occurs earlier, for some value of the phion mass 
$m\equiv m_c$ that is still positive.
 This conclusion is confirmed by the results 
of ref.\cite{rit2} that provides the most accurate non-perturbative
calculation of the 
effective potential of $\lambda\Phi^4$ theories performed so far. 

Understanding the magnitude of $m_c$ requires additional comments. 
The normal ordering prescription in Eq.(1.2) eliminates all ultraviolet 
divergences of the free-field case at $\lambda=0$
but for $\lambda >0$ there are additional divergences. For this reason, one
introduces an ultraviolet cutoff $\Lambda$ and defines the continuum 
theory as a suitable limit $\Lambda \to \infty$.  In this case, however, 
one is faced with a dilemma since a meaningful description of
SSB in quantum field theory {\it must} provide $m_c=0$.
Otherwise, from the existence of a non-vanishing mass gap controlling
the exponential decay of the two-point function of the symmetric phase, and
the basic axioms of quantum field theory \cite{glimm}
 one would deduce the uniqueness of the 
vacuum (and, thus, no SSB). The resolution of this apparent conflict is that 
the continuum limit 
of the cutoff-regulated theory gives a vanishing ratio 
($\epsilon\equiv{{1}\over{\ln {{\Lambda}\over{M_h}} }}$)
\BE
{{m^2_c}\over{M^2_h}} \sim \epsilon 
\EE
so that when the Higgs boson mass $M_h$ is taken as the unit scale of mass,
the possible values of the `phion' mass $0 \leq m \leq m_c$ are
 naturally infinitesimal. 
In this sense, SSB is an
 {\it infinitesimally weak} first-order phase transition where
the magnitude of the ratio
${{m}\over{M_h}}$ represents a measure of the degree of
non-locality of the cutoff-regulated theory.

\par~~~~ii) there is a deep difference between a `free-field' theory and a
`trivial' theory \cite{book} where the interaction effects die out in the
continuum limit.
 The former has a quadratic effective 
potential and a unique ground state. The latter, even for a vanishingly
small strength $\lambda={\cal O}(\epsilon)$ of the elementary two-body 
processes can generate a {\it finite} gain in the energy density, and thus
SSB, due to the macroscopic occupation of the same
quantum state, namely to the phenomenon of Bose condensation. This leads to
a large re-scaling of $\langle \Phi \rangle$. Indeed, one can 
introduce, in general, 
two distinct normalizations for the vacuum field $\phi$, say a `bare' field
$\phi=\phi_B$ and a `renormalized' field
$\phi=\phi_R$. They are defined through the quadratic 
shapes of the effective potential in the symmetric and broken phase 
respectively
\BE
\left. \frac{ d^2 V_{\rm eff}}{d \phi_B^2} \right|_{\phi_B=0} \equiv  m^2, 
\quad \quad \quad 
\left. \frac{ d^2 V_{\rm eff}}{d \phi_R^2} \right|_{\phi_R=v_R} \equiv M_h^2.
\EE
Due to `triviality', the theory is ``nearly'' a massless, free theory so that
 $V_{\rm eff}$ 
is an extremely flat function of $\phi_B$. Therefore, due to (1.5), 
the re-scaling $Z_\phi$ relating $\phi_B$ and $\phi_R$ becomes very large.
 By defining $\phi^2_B = Z_\phi \phi^2_R$, one finds 
$Z_{\phi} ={\cal O}( {{1}\over{\epsilon}} )$ or
\BE
           v_R \sim v_B \sqrt{\epsilon}
\EE
Just for this reason,
 the rescaling of the `condensate' $Z=Z_\phi$ is different 
from the more conventional quantity $Z=Z_{\rm prop}$ defined from the 
residue of the shifted field propagator at $p^2=M^2_h$. According to 
K\"allen-Lehmann decomposition and `triviality' this has
a continuum limit $Z_{\rm prop}=1 +{\cal O}(\epsilon)$.

\par~~~~iii) the existence of
two different continuum limits $Z_\phi \to \infty$ and
$Z_{\rm prop}\to 1$  reflects a fundamental discontinuity in the
2-point function at $p=0$ ($p$= Euclidean 4-vector). This effect is not
totally unexpected and its origin 
should be searched in the infrared divergences of 
perturbation theory for 1PI vertices at zero external momenta \cite{syma}.
Of course, after the Coleman-Weinberg \cite{cw} analysis, we know how to obtain 
infrared-finite expressions for 1PI vertices at zero external momenta. This
involves summing up an infinite series of graphs of different perturbative
order with different numbers of external legs, just as in the analysis of 
the effective potential that was taken as the starting point for our analysis.
In this case, the second derivative of the effective potential gives 
$\Gamma^{(2)}(p=0)$,
the inverse susceptibility
\BE
\chi^{-1}=
\left. \frac{ d^2 V_{\rm eff}}{d \phi_B^2} \right|_{\phi_B=v_B} 
={{M_h^2}\over{Z_\phi}}
\EE
Therefore, if
$Z_{\phi} ={\cal O}( {{1}\over{\epsilon}} )$, one finds
\BE
         {{\Gamma^{(2)}(0)}\over{M^2_h}} \sim \epsilon 
\EE
rather than $\Gamma^{(2)}(0)= M^2_h$ as expected for a free-field theory where
\BE
\Gamma^{(2)}(p) = (p^2 + M^2_h)
\EE
Notice that the discrepancy found in the discrete-symmetry case implies the
same effect for the
zero-momentum susceptibility of the {\it radial} field 
in an O(N) continuous-symmetry theory. This conclusion, besides the
general arguments of \cite{syma}, is supported by the explicit calculations
of Anishetty et al \cite{ani}.

Notice that SSB requires the subtraction of disconnected pieces so that
continuity at $p=0$ does not hold, in general \cite{plb}. At the same time, 
a mismatch at $p=0$ does not violate `triviality' since no scattering 
experiment can be performed with exactly zero-momentum particles.
On the other hand, for large but
finite values of the ultraviolet cutoff $\Lambda$, when `triviality' is not
complete, the discrepancy between 
$\Gamma^{(2)}(0)$ and $M^2_h$ will likely
`spill over' into the low-momentum region 
$p^2 \sim \epsilon M^2_h$. In this region, we expect
sizeable differences from the free-field form Eq.(1.10). 

If really $Z_\phi \neq Z_{\rm prop}$ this result has to show up in
sufficiently precise numerical simulations of the broken phase.
To this end, 
the structure of the two-point function has been probed in refs.
\cite{further} by using the largest lattices considered so far.
 One finds substantial deviations from Eq.(1.10) in the low-$p$ region and
 only for large enough $p$, $\Gamma^{(2)}(p)$ approaches
the free field form (1.10). Also, 
the lattice data of refs.\cite{further}
support the prediction that the 
discrepancy between 
$ \Gamma^{(2)}(0)$ 
and the asymptotic value 
$M^2_h$ becomes larger when approaching
the continuum limit. Notice that 
no such a discrepancy is present in the symmetric phase where 
$\langle \Phi \rangle=0$.

In conclusion: theoretical arguments and numerical evidences suggest that
in the limit $k \to 0$
the excitation spectrum of the broken phase 
can show substantial deviations from the free-field form
$\tilde{E}=\sqrt{k^2+M^2_h}$. Due to the `triviality' of the theory, 
the deviations from the free-field behaviour should, however,
be confined to a range of $ k$ that becomes infinitesimal
in units of $M_h$ in the continuum limit $\Lambda \to \infty$.

\setcounter{equation}{0}
\section{A gap-less mode in the broken phase}

In this section we shall present a simple argument to illustrate
the nature of the excitation spectrum of the broken phase
in the limit $k \to 0$. Our analysis starts from the simple 
relation between the phion density $n$ and the scalar field expectation 
value
\BE
n = \half m \phi^2_B,
\EE
underlying the `particle-gas' picture of ref.\cite{mech}.

Using Eq.(2.1) one can easily transform
the energy density ${\cal E}={\cal E}(n)$ into the effective potential 
$V_{\rm eff}=V_{\rm eff} (\phi_B)$. In this way, 
the $\phi_B=0$ `mass-renormalization' condition
in Eq.(1.6) becomes
\BE
\left. \frac{ \partial {\cal E}}{\partial n} \right|_{n=0} = m.
\EE
Its physical meaning is transparent. If we
consider the symmetric vacuum state 
(``empty box'') and  add a very small density $n$ of phions (each with 
vanishingly small
3-momentum)  the energy density changes by $nm$ in the limit $n \to 0$.
On the other hand, spontaneous symmetry breaking can be viewed as a 
phion-condensation process. This occurs at those values $\phi_B=\pm v_B$ 
where 
\BE
\left. \frac{ d V_{\rm eff}}{d \phi_B} \right|_{\phi_B=v_B}=0
\EE
By using Eq.(2.1) and defining the ground-state particle density
\BE
n_v = \half m v^2_B,
\EE
we also obtain
\BE
\left. \frac{ \partial {\cal E}}{\partial n} \right|_{n=n_v} = 0
\EE
Eq.(2.5) means that small changes of the phion density around 
its stationarity value do not produce any change in the energy of the 
system and
one can add or remove an arbitrary number of
phions at $k=0$ without any energy cost. Just as in the non-relativistic
limit of the theory. Therefore,
the excitation spectrum in the limit $k \to 0$ has no gap and an expansion
in powers of $k$ starts, in general, with a linear term 
$\tilde{E}(k) \sim k$. In this sense, the $k \to 0$
Fourier component of the scalar field, in the broken phase, 
behaves as a {\it massless} field.  We now understand why
the excitation spectrum $\tilde{E}$ 
cannot be $\sqrt{k^2 + M^2_h}$ at low $k$: this 
form does not reproduce $\tilde{E}=0$ for $k=0$. 

Notice that this conclusion, 
although deduced within the framework of ref.\cite{mech}, 
does not depend on the validity of Eq.(2.1). Indeed, Eq.(2.5) follows
from Eq.(2.3) {\it regardless} of the precise functional relation between 
the phion density and the vacuum field. At the same time, this is only possible 
in a first-order phase transition where one can meaningfully investigate
the broken phase in terms of physical elementary 
excitations of the symmetric phase.
In this case in fact, however small the phion mass $m$ can be, there exists
a non-relativistic limit $k \ll m$ where the scalar
condensate will respond in a phase-coherent way. In this sense, 
the gap-less mode can be considered 
the Goldstone boson of a spontaneously broken {\it continuous}
symmetry, the phase rotations of the condensate wave-function, 
that does not exist in the symmetric phase.

After this general discussion, let us now attempt
a quantitative description of the energy spectrum of the broken phase.
A first observation is that  
for large enough $k$ we expect
\BE
\tilde{E}(k) \sim \sqrt{k^2 +M^2_h}
\EE
On the other hand, the region $k \to 0$ 
of  {\it low-density} Bose systems
can be analyzed in a universal way \cite{trento} namely
\BE
     \tilde{E} \sim c_s k~~~~~~{\rm for~k\to 0}
\EE
where $c_s$ is the sound velocity 
\BE
            c_s \equiv {{1}\over{m}} \sqrt{4\pi n_v a}
\EE
Here $a\sim {{\lambda}\over{8\pi m}} $ 
denotes the S-wave `phion-phion' scattering length which enters the expression
for the Higgs mass \cite{mech}
\BE
M^2_h \equiv 8\pi n_v a
\EE
Notice that Eq. (2.7) 
becomes a better and better approximation 
in the limit of very low-densities
$n_va^3 \to 0$ where all condensed phions are found in the state
at $k=0$ and there is no population of the finite momentum modes (`depletion').

Let us analyze the situation in the case of 
spontaneous symmetry breaking in cutoff $\lambda\Phi^4$ theory. On one hand,
`triviality' requires a continuum limit with
 a vanishing strength $\lambda={\cal O}(\epsilon)$ 
for the elementary 2-body processes. On the other hand, 
together with Eq.(1.5), this leads to $aM_h \sim \sqrt{\epsilon}$. Therefore, 
 since one takes $M^2_h\equiv 8\pi n_v a$ 
as  a cutoff-independent quantity, we find
\BE
                   n_v a^3 ={\cal O}(\epsilon)
\EE
When $\epsilon \to 0$,
the phion-condensate becomes infinitely dilute so that
the average spacing between two phions in the condensate, 
$d \equiv n_v^{-1/3}$ becomes enormously larger than their scattering length.
In this limit, 
the energy spectrum (2.7) becomes exact ( in the limit $k \to 0$). 

Notice, however, that the phion density
$n_v={{1}\over{2}}m v^2_B$, is very large, 
${\cal O}(\epsilon^{-1/2})$, in the physical units denoted by the correlation
length $\xi_h\equiv1/M_h$. Indeed, 
${{d}\over{\xi_h}} \sim \epsilon^{1/6}$.  It is because there is such a 
high density of phions that their tiny 2-body
interactions ${\cal O}(\epsilon)$ can produce a finite effect 
on the energy density.

In conclusion: spontaneous symmetry breaking in a cutoff $\lambda\Phi^4$ theory
gives rise to an excitation spectrum that is {\it not} exactly 
Lorentz-covariant. The usual assumption 
$\tilde{E}(k) \sim \sqrt{k^2+M^2_h}$ 
is not valid in the limit $k \to 0$ where one actually finds a `sound-wave' 
shape $\tilde{E}(k) \sim c_s k$. This result reflects the
{\it physical} presence of the scalar condensate.

\setcounter{equation}{0}
\section{A long-range potential in Higgs condensates}

It is well known that
condensed matter systems can support long-range forces
even if the elementary constituents have only short-range 2-body 
interactions. Just for this reason, it is not surprising that
the existence of a gap-less mode for $k \to 0$ in the broken phase
can give rise to a long-range potential. For instance, when
coupling fermions to a (real) scalar Higgs field with vacuum 
expectation value $v$ through the Standard Model interaction term 
\BE
         -m_i \bar{\psi}_i \psi_i (1+ {{h(x)}\over{v}})
\EE
the static limit $\omega \to 0$ of the Higgs propagator
\BE
                D(k,\omega)= {{1}\over{\tilde{E}^2(k)- \omega^2 -i0^+}}
\EE
gives rise to an attractive potential
between any pair of masses $m_i$ amd $m_j$ 
\BE
          U(r)=-{{m_im_j}\over{v^2}} 
\int {{d^3k}\over{(2\pi)^3}} 
{{ \exp (i {\bf{k}}\cdot {\bf{r}})  } \over{  \tilde {E}^2(k)   }}
\EE
By assuming Eqs.(2.6) and (2.7) for $ k \to \infty$ and $k \to 0$, 
and using
the Riemann-Lebesgue theorem \cite{goldberg} on Fourier transforms, the
leading $r \to \infty$
behaviour is universal. Any form of the spectrum that for $k \sim m$
interpolates between the two asymptotic trends would produce the same
result. At large distances $r >> 1/m$ one finds
($\eta \equiv c^2_s={{M^2_h}\over{2m^2}}= {\cal O} ({{1}\over{\epsilon}})$)
\BE
            U(r)=- {{G_F}\over{4\pi \eta}}{{m_im_j}\over{r}}
[1 + {\cal O}(1/mr)]
\EE
where $G_F\equiv 1/(v^2)$. 
In the physical case of the Standard Model one would identify
$G_F \sim 1.1664 \cdot 10^{-5}$ GeV$^{-2}$ with the Fermi constant.
Notice that the coupling in Eq.(3.1) naturally defines the `Higgs charge' 
of a given fermion as its physical mass. However, 
for nucleons, this originates from
more elementary Higgs-quark and Higgs-gluon interactions. These effects 
can be resummed to all orders by replacing Eq.(3.1) with the alternative 
expression through the trace of the energy-momentum tensor
$\theta^{\mu}_{\mu}$, namely
\BE
         - \theta^{\mu}_{\mu} (1+ {{h(x)}\over{v}})
\EE
This different normalization of the Higgs-fermion coupling reduces to the
usual definition in the case of free quarks and yields exactly the nucleon mass
when evaluating the matrix element between nucleon states 
(see for instance \cite{okun} )
\BE
        \langle N | \theta^{\mu}_{\mu}| N \rangle =m_N \bar{\psi}_N \psi_N
\EE
Notice that Eq.(3.5) is formally analogous to a Brans-Dicke theory \cite{brans}.
Here, however, the framework is very different since the $h-$field propagates
in the presence of the phion condensate. Also, Eq.(3.5) 
represents the appropriate form to implement a `strong 
form' of the Equivalence Principle \cite{weinberg} by extending the notion of
inertia to the electromagnetic field through scale 
non-invariant quantum effects \cite{collins}. 

We note that the strength of the long-range potential is proportional to the
product of the masses and is
naturally infinitesimal in units of $G_F$. It would vanish 
in a true continuum theory. 
Therefore, it is natural to relate this extremely weak
interaction to the gravitational potential
and to the Newton  constant $G$ by identifying
\BE
              \sqrt{\eta} = \sqrt{ {{G_F}\over{G}} } \sim 10^{17}
\EE
Notice, that the long-range $1/r$ potential is a direct consequence 
of the existence of the scalar condensate. Therefore, speaking of
gravitational interactions makes sense only for particles that can 
induce variations of the phion density by exciting the gap-less
mode of the Higgs field.
 In this sense, phions, although possessing an inertial mass, have
no `gravitational mass'. 

In the physical case of the Standard Model, and assuming the
range of Higgs mass
$M_h \sim 10^2-10^3$ GeV, this  leads to 
a range of phion masses
$m \sim 10^{-4}-10^{-5}$ eV. 
The detailed knowledge of the spectrum 
$\tilde{E}(k)$ for $k\sim m $ would allow to compute the terms
${\cal O}(1/mr)$ in Eq.(3.4) and predict a characteristic pattern of
 `fifth force' deviations below the centimeter scale.

\setcounter{equation}{0}
\section{Summary and concluding remarks}

In this paper we have presented a simple physical picture where
Newtonian gravity can be interpreted 
as a long-wavelength excitation of the scalar
condensate inducing spontaneous symmetry breaking.
 We emphasize that our
main result in Eq.(3.4) depends only on the Riemann-Lebesgue theorem on Fourier 
transforms \cite{goldberg} and 
two very general properties of the excitation spectrum. Namely, the
 `diluteness'  condition Eq.(2.10) (that leads to the `sound-wave' shape in
Eqs.(2.7) and (2.8) for $k \to 0$) and
the Lorentz-covariance for large $k$ (that leads to Eq.(2.6)). 
These general properties are expected to occur in
{\it any} description of spontaneous symmetry
breaking in terms of a weakly coupled Bose field. We emphasize, as in sect.2, 
that the assumption of a weakly first-order phase transition is essential.
Only in this case, in fact, there is a non-relativistic regime $k \ll m$ where
the scalar condensate reacts with phase-coherence \cite{vector}.
A more complete
description of gravitational phenomena requires the detailed form of the
spectrum $\tilde{E}(k)$ and, in particular, the precise knowledge
of the phion mass $m$.  Deviations from the Newton potential are expected
at typical distances $r\sim 1/m$ and could, eventually, be
detected in the next generation of precise `fifth-force' experiments 
\cite{price}. 

We stress that
the apparently `trivial' nature of $\lambda\Phi^4$ theories in four 
space-time dimensions should not induce to overlook the possibility
that gravity can arise as a gap-less mode of the Higgs field.
Indeed, our description is only possible if one assumes the existence of 
an ultimate ultraviolet cutoff so that the natural formulation of the
theory is on the lattice and `triviality' is never complete.
In this case, however, the existence of a non-trivial infrared behaviour 
in the broken phase
can be guessed from the equivalence of low-temperature Ising models with highly
non-local membrane models on the dual lattice \cite{gliozzi} whose continuous
version is the Kalb-Ramond \cite{kalb}
model. Thus, in the end, a Higgs-like description of gravity
 would turn out to be equivalent, at some scale, to a 
Feynman-Wheeler theory of {\it strings}, 
as electromagnetism for point particles. 
This may be useful to establish a link between our description and 
alternative pictures of gravity, even with very different degrees of
locality recalling, however, that the existence of the cutoff cannot be
neglected (for instance when comparing with the action-at-distance theory
of Hoyle and Narlikar \cite{hoyle}). At the same time, the basic idea that
one deals with the {\it same} theory should allow to replace
a description with its `dual' picture when better suited to provide
an intuitive physical insight. 

Many readers, we realize, will be reluctant since
the presently accepted point of view tends to regard
Newtonian gravity as a  well defined limit of a more fundamental theory, 
namely Einstein gravity. This is 
believed to lie outside of the Standard Model and 
to require fundamental and genuinely new interactions and particles (spin-2 
gravitons). Apparently, 
this point of view is very natural since Einstein
equations in the weak-field limit
can be obtained \cite{weinberg2}
from flat space by requiring the Lorentz invariance
of matrix elements for absorbing and emitting massless spin-2 quanta. 
However, this derivation assumes the validity of 
a fully Lorentz-covariant description of the elementary gravitational processes. 
In the presence of spontaneous symmetry breaking, this assumption is not 
necessarily true in the limit $k \to 0$ if the
departure from an exact Lorentz-covariant spectrum 
is the origin of gravitational interactions (and a consequence of
the cutoff). 

Quite independently of any application to gravity, the possible 
departure from a Lorentz-covariant excitation spectrum
is a general feature in 4-dimensional
interacting theories \cite{segal}.
 The reason is that the usual normal-ordering procedure guarantees the 
local commutativity of Wick-ordered products of the field operator in the free
theory but no such a procedure is known {\it a priori} for the interacting case.
Thus the argument is circular since the proper
normal-ordering procedure is only known after
determining the vacuum and its excitation spectrum. In this respect, the
usual approach, where one
uses the normal-ordering definition of
the free-field theory and defines the continuum theory as a suitable limit 
$\Lambda \to \infty$, {\it is} consistent. In fact, in this limit, the 
spectrum reduces simply to $\sqrt{k^2 + M^2_h}$ for all $k$ (except $k=0$).

On the other hand, without considering these technical details, 
the very accurate equality between the inertial and
gravitational mass of known particles
should convince a skeptical reader that our result is not
totally unexpected. Actually, to a closer inspection, a tight link between 
the physical origin of gravity and the physical
origin of inertia is unavoidable. Only in this 
case, in fact, one can fully understand
what, after Einstein, it is called 
Mach's Principle, namely the consistent
vanishing of inertia if gravity would be switched off \cite{pauli}. 
The point is that Einstein's description of gravity is purely geometric and 
macroscopic. As such, it does not depend on any hypothesis about
the  physical origin of this interaction. Einstein's construction would remain
the same if the Newton potential would be {\it experimentally} known to behave
as ${{1}\over{r}} \exp(-\mu r)$ rather than as $1/r$. Just for this reason, 
 general relativity, by itself, is unable to
 predict \cite{pauli} even the {\it sign} of the gravitational force 
(attraction rather than gravitational repulsion). Rather, Einstein had to 
start from the peculiar properties of
Newtonian gravity to get the basic idea to transform the classical effects of
this type of
interaction into a metric structure. 
In this sense, it is not surprising that a few `crucial tests' of general 
relativity in weak gravitational field merely verify
\cite{schiff} two well established 
structures, namely Special Relativity (SR) and the Equivalence Principle (EP), and
do not necessarily require a fundamental tensor theory. The reason is
that the infinitesimal transformation to the rest frame
of a freely falling elevator is of very general nature. For instance, it
can also be obtained with a conformal transformation of length, time and 
mass \cite{fulton}. Just to illustrate this point, let us consider the
relation among reference frames
in the gravitational field of a large 
mass $M$ (e.g. the sun). Up to higher powers of the gravitational strengths, any
bound observer $O(i)$ can be considered as performing circular orbits of radius
$r(i)$.  The ordering of the observers 
is such that $r(i)< r(i+1)$ so that, assuming an overall
weak-field condition $ {{2 GM}\over{c^2r(1)}} \ll 1$,  
the relation to the asymptotic reference frame $K(0)$ at spatial 
infinity can be approximated as an infinitesimal Lorentz-transformation 
with the radial `escape' velocity
\BE
           v^2(i)= {{2 GM}\over{r(i)}}
\EE
The set of metrics
\BE
      ds^2(i)= c^2dt^2(i) - dr^2(i) -
r^2(i)\left[d\theta^2(i)+ \sin^2\theta(i)d\varphi^2(i)\right]
\EE
for the $O(i)$ frames implies
\BE
      ds^2(0)= c^2dt^2(i) \left[1- {{2 GM}\over{c^2r(i)}}\right] -
{{dr^2(i)} \over{1 - {{2 GM}\over{c^2r(i)}} } } -
r^2(i)\left[d\theta^2(i)+ \sin^2\theta(i)d\varphi^2(i)\right]
\EE
for the $K(0)$ frame that, 
 indeed, is the Schwarzschild metric. The weak-field restriction means that
Eq.(4.1) is valid up to higher order terms. For instance, one could replace 
the relativistic expression for the kinetic energy and obtain
\BE
      ds^2(0)= {{c^2dt^2(i)}\over { \left[1+ {{ GM}\over{c^2r(i)}}\right]^2}} -
dr^2(i) \left[ 1 + {{ GM}\over{c^2r(i)}} \right]^2  -
r^2(i)\left[d\theta^2(i)+ \sin^2\theta(i)d\varphi^2(i)\right]
\EE
In this respect, Einstein's definition of the Mach's Principle
provides the real physical meaning of his description of gravity. Namely, 
in a theory where the common origin of inertia and gravity is 
built in, general relativity can represent a very
elegant and clever way to compute the weak-field
corrections to measurements of length and time. After all, 
this interpretation of general relativity is consistent with its 
name that suggests a description of relative effects in gravitational fields
rather than  a truly dynamical explanation of the origin of the gravitational
force. This interpretation 
is also consistent with alternative views \cite{fuji,zee,adler}
that consider classical Einstein gravity
as a weak-field {\it effective} theory generated by
underlying quantum phenomena.

The new result is that
the crucial ingredient for the Equivalence Principle, namely
`inertial mass=gravitational mass', turns out to be a
deducible consequence of our present understanding of electroweak interactions
where the non-trivial coexistence of `SR' and `EP'  does not 
require to change the space-time but depends on the particular nature
of the ground state.
This also removes all self-contradictory consequences of accomodating
gravity within exact Lorentz-covariance \cite{misner} and
represents a step towards 
a more comprehensive theory whose final form, of course, 
we are not able to predict. 

A definite prediction, however, is that the gravitational force
is naturally instantaneous.
The velocity of light $c$ has nothing to do with 
the long-wavelength excitations of the phion condensate that for $k \to 0$
propagate with the fantastically high speed 
$c_s= \sqrt{\eta} c\sim 10^{17} c$.
On the other hand, for $k \sim m$, i.e. at the joining of 
the two branches of the excitation spectrum, one recovers the expected result 
$d\tilde{E}/dk < c$.  To a closer inspection, 
this apparently bizzarre result appears
less paradoxical than the generally accepted point of view that considers
the inertial forces in an accelerated laboratory as the consequence of a 
gravitational wave generated by distant accelerated matter. Indeed, if the
gravitational interaction propagates with the light velocity, distant matter
must be accelerated {\it before} the inertial 
reaction is actually needed \cite{dicke}. The 
same type of conclusions is suggested by the analysis of tideal forces 
\cite{narlikar}. 

Notice that, quite independently of our results and 
within the generally accepted framework of general relativity, the existence
(or not) of faster than light signals is still an open  possibility.
For instance, regardless of the quantum phenomena that give rise to the 
ground state, it is known since a long time
that constant energy-density solutions
of Einstein equations contain, indeed, closed time-like curves \cite{godel}.
More recently, the same type of questions have been raised by
`inflation' that, indeed, represents a superluminal mode
of expansion introduced 
 to reconcile the predictions of relativistic cosmology with the
size of the observed universe. In the present approach \cite{guth}, 
this is achieved through an {\it extremely flat}
scalar potential that is very natural 
in a description where spontaneous symmetry 
breaking is an {\it infinitesimally weak} first-order phase transition. 
Actually, just the possibility to understand the extraordinary 
fine-tuning needed in inflationary models was one of the motivations
behind the early attempt \cite{mech}
to explain inertia and gravity from the same physical phenomenon. In fact, 
due to the large re-scaling $Z_\phi \sim \eta$ 
in Eqs.(1.6) and (1.7), one has actually a `2-parameter' theory where 
the `renormalized' value $v_R \sim G^{-1/2}_F$ is used for the W mass 
$M^2_w \sim {{g^2v^2_R}\over{4}}$ and the `bare' value 
$v_B \sim G^{-1/2}$ is used to generate Einstein's lagrangian 
${\cal L}\sim Rv^2_B$. After all, taking into account the crucial quantum 
phenomenon of the vacuum field rescaling, this is in the spirit of a
spontaneously broken theory of {\it classical} Einstein 
gravity \cite{fuji,zee,adler} where gravity, however, is induced by
the Standard Model Higgs field \cite{bij,bij2}. After some thought, 
however, this picture can hardly work. Indeed, besides introducing 
non local effects that is difficult to accomodate within the traditional
Einstein picture, 
the non-trivial presence of the phion condensate would likely require
the introduction of a preferred metric structure. In fact,
this unusual form of matter has an inertial mass and transmits gravity
but does not generate any curvature: it is the quantum realization of the 
old-fashioned {\it weightless} aether whose long-range
density oscillations are determined by the coupling of known matter to the 
gap-less mode of the Higgs field.

To better clarify this point, and in the spirit of a weak-field analogy,
we note that in classical
Einstein theory, all forms of energy and matter contribute to the 
space-time curvature. Therefore, to take into account
the peculiar medium that transmits
gravity, one has to give up full general covariance.
One possibility would be to require 
$\sqrt{g}=1$ \cite{vandam,wilc}
so that Einstein equations are replaced by their traceless counterpart.
In this constrained formulation of gravity, there is no cosmological term
from spontaneous symmetry breaking and, whatever the distribution of the known
forms of matter,
the trace of the energy-momentum tensor coincides with (minus)
the scalar curvature $R$, up to an integration constant \cite{wilc}.
 Therefore, to describe classical motions 
and up to higher-order $O(G^2)$ and $O(G G_F)$ terms, 
Eq.(3.5) can be used to
generate Einstein lagrangian in a $\sqrt{g}=1$ world. 
 Connection with other descriptions of
gravity (again for $\sqrt{g}=1$) can be established if we
separate out in Eq.(3.5)
the long-wavelength part of the Higgs field  $\tilde{h}(x)$
associated with the linear excitation
spectrum of the scalar condensate. In this situation, and
up to higher order terms, 
one can replace the lagrangian in Eq. (3.5) with 
($\phi(x)\equiv {{\tilde{h}(x)}\over{v}}$)
\BE
            {{1}\over{16 \pi G}} \exp(\phi) R 
\EE
Eq.(4.5), formally, rensembles string-inspired  descriptions 
of Einstein gravity where, however, the `dilaton' is identified with
the gap-less, `tachionic' mode of the Higgs field. In this respect, we find
surprising that theoretical models of string cosmology seem
to indicate, indeed,
 the `tachionic' nature of the dilaton \cite{venezia}, at least
in an early epoch of inflation where space-time is nearly flat. 
As anticipated, this may
be a consequence of the non-trivial duality properties \cite{gliozzi} 
of four-dimensional broken-symmetry Higgs phases. 

To conclude, we want to mention two possible crucial implications of the 
scalar condensate at the cosmological level:
\par~~~a)
our picture, despite of its conceptual simplicity,
provides a natural solution of the
so-called `hierarchy-problem'. This depends on 
the {\it infinitesimally weak} first-order nature of the phase
transition: in units of the Fermi scale, 
 $M_h\sim G_F^{-1/2}$, the Planck 
scale $G^{-1/2}$ would diverge for a vanishing phion mass $m$. 
These three scales are hierarchically 
related through the large number $\sqrt{\eta} \sim 10^{17}$ that is the only
manifestation of an ultimate ultraviolet cutoff. In this sense, spontaneous
symmetry breaking in $(\lambda\Phi^4)_4$ theory represents an
approximately scale-invariant 
phenomenon and it is conceivable that powers of
the `replica-factor' $10^{17}$ will further show up in a natural way.
This speculation may not be too far from
the actual physical situation in large-scale astronomy \cite{rowan} where the
strong experimental evidence
for a hierarchical cosmology was clearly pointed out by 
de Vaucouleurs \cite{devau}
long time ago. As a consequence, by
accepting the common physical origin of
both inertia and gravitation and the crucial role of the scalar condensate, 
it becomes natural to search the origin of 
the observed self-similarity of the universe \cite{sylos} 
in the basic features of spontaneous symmetry breaking. In this way, one can
hope to resolve the serious discrepancy \cite{sylos}
 between the present models of galaxy formation 
and the present models of expansion of the universe.

~~~b) as pointed out in sect.2, one expects
tiny departures  from the linear excitation
spectrum at small $k$ due to residual self-interaction effects within the 
condensate. The deviations from an exact `superfluid-regime' can  introduce a
small friction that could become important in the typical astronomical
large-scale
and low-acceleration conditions. These effects, totally unobservable in 
laboratory tests, would show up
as small deviations from the Newton force and could be interpreted
as modifications of inertia and/or of gravity. It is not
unconceivable that a closer look to the small residual self-interactions in
the scalar condensate can explain 
the peculiar modification of inertia suggested by Milgrom \cite{milgrom}
to resolve the observed
mass discrepancy in many galactic system. Indeed, there is a
very close analogy between the typical non-linear effects associated with the 
response of a non-trivial vacuum \cite{piran} and those actually needed to
explain this particular effect \cite{milgrom2}.
Experimentally, this shows up
when the acceleration of bodies becomes comparable to a universal acceleration
field 
$a_o\sim 10^{-8}~cm~sec^{-2}$. Remarkably, this value is of the same
order as (the unexplained part of) the anomalous acceleration toward the sun
suggested by the Pioneer 10/11
\cite{pioneer} data. In our picture of gravity, it would be natural to interpret
such effects as a tiny  `braking' of the scalar condensate.

\vskip 20 pt
{\bf Acknowledgements}~
We thank M. Baldo, P. Cea, L. Cosmai, P. Dalpiaz, J. Hosek, 
P. M. Stevenson and J. J. van der Bij for useful
discussions.


\begin{thebibliography} {99}
\bibitem{vector}
Notice that the phase
transition remains first-order if spontaneous symmetry breaking is induced by
(or contains the additional contributions of) intermediate vector bosons 
\cite{cw,ibanez}.
\end{thebibliography}
\end{document}